\documentclass[]{aa}
\usepackage{graphicx}
\usepackage{amssymb}
\usepackage{setspace}
\begin{document}

\title{Determination of the size, mass, and density of ``exomoons'' from photometric transit timing
variations}

\titlerunning{Exomoons in transiting systems }

\author{
	A. Simon\inst{1},
	K. Szatm\'ary\inst{1} \and
	Gy. M. Szab\'o\inst{1,2}
	}

\authorrunning{Simon et al.}

\offprints{asimon@titan.physx.u-szeged.hu}

\institute{
	\inst{1} Department of Experimental Physics \& Astronomical Observatory,
	University of Szeged,
	6720 Szeged, Hungary\\
	\inst{2} Magyary Zolt\'an Research Fellow\\
	\email{asimon@titan.physx.u-szeged.hu}
	}

\date{}

\abstract
	{}
	{Precise photometric measurements of the upcoming 
	space missions allow the size, mass, and density of satellites of exoplanets to be determined.
	Here we present such an analysis using the photometric transit timing variation (${\rm TTV_p}$).
	}
	{
	We examined the light curve effects of both the transiting planet and its satellite. 
	We define the photometric central time of the transit that is equivalent to the transit of 
	a fixed photocenter. This point orbits the barycenter, and leads to
	the photometric transit timing variations.
	}
	{
	The exact value of  ${\rm TTV_p}$
	depends on the ratio of the density, the mass, and the size of the satellite
	and the planet. Since two of those parameters are independent, a reliable estimation 
	of the density ratio leads to an estimation of the size and the mass of the exomoon. 
	Upper estimations of the parameters are possible in the case when an upper limit of ${\rm TTV_p}$ is known.
	In case the density ratio cannot be estimated reliably, we propose an approximation with assuming
	equal densities.
	The presented photocenter ${\rm TTV_p}$ analysis predicts the size of the satellite better than the mass. 
	We simulated transits of the Earth-Moon system in front of the Sun.
	The estimated size and mass of the Moon are 0.020 Earth-mass and 0.274 Earth-size if equal
	densities are assumed. This result is comparable to the real values within a
	factor of 2. If we include the real density ratio (about 0.6), the results are 0.010 Earth-Mass
	and 0.253 Earth-size, which agree with the real values within 20\%{}.
	}
	{}


\maketitle


\section{		Introduction}

The discovery of planets in other solar systems (``exoplanets'') 
led to the question of whether these planets also have satellites (``exomoons'').
The photometric detection of such a satellite was first 
suggested by Sartoretti \& Schneider (1999, SS99 in the followings; also discussed by Deeg, 2002),
who gave a method for estimating the mass of this exomoon. The idea is to calculate the
baricentric transit timing variations (${\rm TTV_b}$).
{ This quantity has been applied to the upper mass estimations of hypothetical satellites around transiting
exoplanets (Charbonneau et al. 2005; Bakos et al. 2006; McCullough et al. 2006; Steffen et al. 2005;
Gillon et al. 2006).}
In this theory, the barycenter orbits the star with a constant velocity, and it 
transits strictly  periodically.
As the planet (and the satellite, too) revolves around the planet-satellite 
barycenter, the transit of the planet wobbles in time. The
${\rm TTV_b}$ is based on finding the center between ingress/egress or from some other method known from
eclipse minimum timing. The circular velocity 
around the star can be expressed via the $m_*$ mass of the star and the $a_p$ semi-major axis
of the planet as $v=\sqrt{\gamma m_*\over a_{p}}.$ 
The maximal normal amplitude (i.e. half of the peak-to-peak amplitude) is finally 
\begin{equation}
	{\rm TTV_b} = {1\over v}B_c, 
\end{equation}
where $B_c={a_sm_s\over m_p+m_s}$ is the distance from the centerpoint of the planet to the barycenter.
Here $a_s$ is the semi-major axis of the satellite, and $m_s < m_p$ are the mass of
the satellite and the planet, respectively. If ${\rm TTV_b}$ is observed or if there is an upper limit,
the mass is estimated by setting $a_s$ to the borders of Hill-stable regions, 
$a_s=a_H \equiv a_p \left( {m_p\over 3m_*} \right)^{1/3}$ the Hill-radius. 
With this selection, $m_s$ remains the single parameter and it can be determined as
\begin{equation} 
m_{s,SS99}=
{v\ {\rm TTV_b} \over a_s} m_p= {v \ {\rm TTV_b} \over a_p} \left( 3m_p^2 m_* \right)^{1/3}
\end{equation}
(Eq. (24) in SS99).
This model can be refined by including the results of Barnes and O'Brien (2002),
who suggests more stringent constraints on the survivability of exomoons: 
the maximal distance from the planet
is about $a_H/3$, depending slightly on the mass ratio. More recently, Domingos et al. (2006)
have suggested that the maximum semimajor axis is $0.4895 a_H$ for a prograde satellite
and is $0.9309 a_H$ for a retrograde satellite, both in circular orbit and with a mass ratio
$m_s/m_p=0.001$. With the
realistic choice of $a_s=a_H/3$, the satellite
masses would lead to better estimates that are about three times larger than given by Eq. 2. 
However, this modification has been rarely adopted in the literature.

In Szab\'o et al (2006) we designed numerical simulations to predict how probable the 
discovery of an exomoon is with a photometric technique. For this detection
we suggested using the (photometric) central time of the transit, $\tau$ as
\begin{equation}
\tau={\sum t_i \Delta m_i \over \sum \Delta m_i},
\end{equation}
where $m_i$ is the $i$-th magnitude measurement at $t_i$ time.
We show later the {\it photometric transit timing variation}, $\rm TTV_p$ calculated from this $\tau$
{ requires its own physical interpretation, which differs from} the conventional $\rm TTV_b$. 
The difference is to account for the 
photometric effects of {\it both} the planet and the satellite:
even when the satellite cannot be detected directly in the light curve, 
its presence can actually dominate ${\rm TTV_p}$.  
We have shown that $\tau$ is robust, therefore it is applicable even in the case of as ``noisy'' data
series as the COROT and Kepler missions provide. These missions are able to detect a Moon-like satellite 
around an Earth-like planet with 20\% probability (Szab\'o et al., 2006).

In this paper we give the complete analytical description of this theory of photometric timings,
directly leading to 
estimation of various parameters such as the mass and the radius of the satellite. 
We test the formulae by modeling the Earth-Moon system.

\section{		Light curve morphology}

Let us assume a planet and its satellite transiting the star. The position of the planet and the
satellite is such that their images do not overlap as seen by the observer.
{ We then assume that the relative position of the planet and the satellite 
does not change during the transit;
the transit is central and the orbital inclination of the satellite is 0.}
Let the sampling of the light curve be uniform.
Under these assumptions, $\tau$ is exactly the barycenter of the light curve 
(i.e. considered as a polygon). 
Furthermore, let 
\begin{equation}
	\mu=m_s/m_p,\ \ \  \ \chi=\rho_s/\rho_p,\ \ \  \ \vartheta=r_s/r_p
\end{equation}
be the ratio of the masses, densities and sizes (radii), respectively. 
{ Any of the three} can be expressed from the two other ones. In the case of the direct detection
of the satellite in the shape of the light curve, 
one can directly calculate $\vartheta^2=\Delta m_s/\Delta m_p$, where $\Delta m_s$ and $\Delta m_p$ 
are the brightness { decreases} due to the satellite and the planet, respectively. 
But even if the noise level is too high for a direct detection, the presence of a moon can be reasoned indirectly
(Szab\'o et al. 2006), as formulated in the following.

The shape of the light curve is the sum of two components: a single planet
and a single moon in transit (Fig. 1). 
The shape of those components is
\begin{equation}
\Delta m_{s,ti} = {r_s^2 \over r_*^2} \ f(t_i-\tau_s),
\end{equation}
\begin{equation}
\Delta m_{p,ti} = {r_p^2 \over r_*^2} \ f(t_i-\tau_p),
\end{equation}
where $\Delta m_{s,ti}$ and $\Delta m_{p,ti}$ are the magnitude decreases at
time $t_i$, and $\tau_p$ and $\tau_s$ are the {times when the planet 
and the satellite each passes alone before the central meridian of the star. 
Here $f(x)$ is  the normalized shape function of the transit.}
We can assume that $f(x)$ is an { axially} symmetric light curve to
$x=0$, { and its off-transit value is} $0$.

If we consider the transit of the planet and the moon together,
\begin{equation}
\Delta m_i=\Delta m_{s,t_i}+\Delta m_{p,t_i}.
\end{equation}
In this case we know from Eqs. 3 and 7 that
\begin{equation}
\tau={ \sum t_i \Delta m_{p,ti} + \sum t_i \Delta m_{s,ti} \over \sum\Delta m_i}.
\end{equation}

\begin{figure}
\includegraphics[width=9cm]{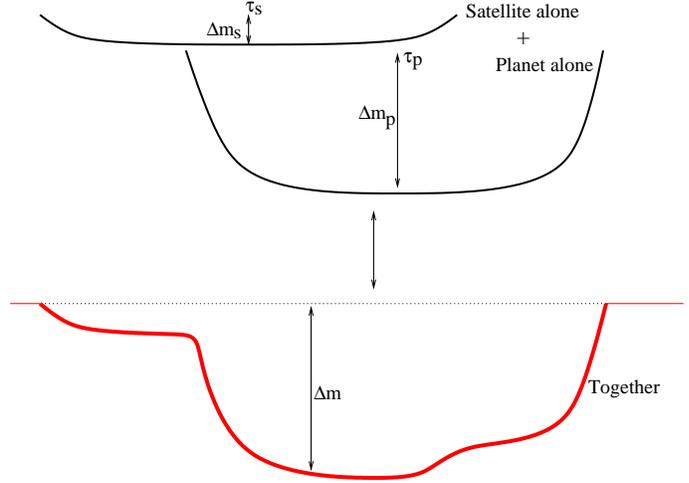}
\centering\caption{Composition of the transit light curve from $\Delta m_p$ and $\Delta m_s$.}
\end{figure}
Let $N=\sum f(t_i)$ be the ``area'' under the function $f$. 
With this, we can easily rewrite the nominator as
\begin{equation}
\sum\Delta m_i={r_s^2+r_p^2 \over r_*^2} N.
\end{equation}
Taking the axial symmetry of $f$ into account, the numerator of Eq. 8 can be similarly rewritten as
\begin{equation}
\sum t_i \Delta m_{s,ti} + \sum t_i \Delta m_{p,ti}=
{r_s^2\over r_*^2}N\cdot \tau_s+
{r_p^2\over r_*^2}N\cdot \tau_p ,
\end{equation}
and finally
\begin{equation}
\tau={ {r_s^2\over r_*^2}N\cdot \tau_s+{r_p^2\over r_*^2}N\cdot \tau_p \over 
{r_s^2+r_p^2 \over r_*^2} N}={r_s^2\tau_s+r_p^2\tau_p \over r_s^2+r_p^2}.
\end{equation}

Now let us consider an extreme { configuration} with a leading moon.
{ We consider the first contact time, when the ingress of the satellite 
begins, and relate it to the time when the satellite is exactly
in the center of the stellar disc.
Between these events, $\tau_s={1\over v} (r_s+{r_*})$ time has to elapse.}
The time between the first contact of the satellite and the time when the planet is in front of the center of the stellat disk
is $\tau_p={1\over v} (r_s+{r_*}+a_s)$.
Substituting these and the analogous formulae into Eq. 11,
\begin{equation}
\tau={1\over v} \left( r_s + {a_s r_p^2 \over r_s^2+r_p^2}+{r_*}\right) 
\end{equation}
with a leading satellite and
\begin{equation}
\tau={1\over v} \left( r_p + {a_s r_s^2 \over r_s^2+r_p^2}+{r_*}\right) 
\end{equation}
with a trailing satellite.

The { basic} conclusion is that the { planet-satellite system can be substituted
by a single celestial body, which can cause the same $\rm TTV_p$ and}
{ which} is located on the planet-moon line at 
$P_c = a_s {r_s^2\over r_s^2+r_p^2}$ distance
from the planet. We call this { particular} point as {\it ``photocenter''} in the following.

The barycenter lies at $B_c={a_s m_s \over m_s+m_p}$ distance from the planet and it revolves strictly periodically.
The photocenter lies somewhere else between the planet and the satellite, therefore the $\tau$ time of the transit
will wobble in time (See Fig. 2). The time between
the transit of the barycenter and the photocenter is $\rm TTV_p$, which can be expressed as
\begin{eqnarray} \begin{array}{r}
	{\rm TTV_p}={1\over v}\left| P_c-B_c\right| ={1\over v}\left| {a_sr_s^2\over r_p^2+r_s^2} - {a_sm_s\over
	m_p+m_s}\right|=\\
	{a_s\over v}\left| {\vartheta^2\over 1+\vartheta^2} - {\mu\over 1+\mu}\right|.
\end{array}
\end{eqnarray}

\begin{figure}
	\includegraphics[width=\columnwidth]{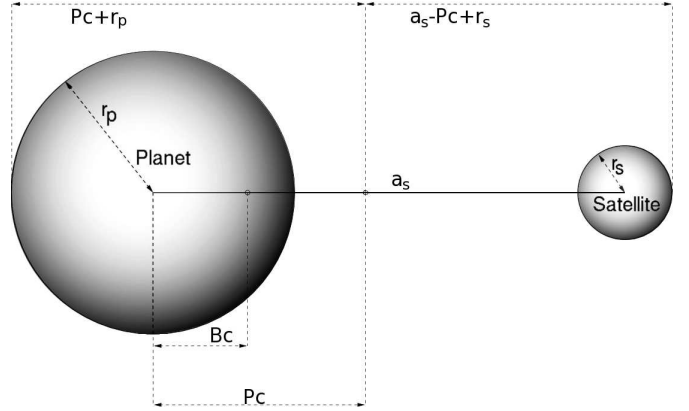}
	\caption{Position of the barycenter and the photocenter in a transiting system (not to scale).}
\end{figure}

With neglecting $\vartheta^2$ and $\mu$ to 1 ($\vartheta^2\ll 1$, $\mu\ll 1$, 
their values $\lessapprox 0.01 $) in the nominators, ${\rm TTV_p}$ simplifies to our basic equation
\begin{equation}
	{\rm TTV_p} = {a_s\over v}\left|\vartheta^2-\mu\right|,
	\label{S1}
\end{equation}
where $\mu=m_p/m_s$ is the mass ratio and $\vartheta=r_s/r_p$ the ratio of radii.
The comparison of Eqs. 1 and 14 shows the difference between the SS99-theory and the present approach:
the introduction of the $P_c$ term. 
This reflects the { different basic concepts.
In the SS99-theory, ${\rm TTV_b}$ is a dynamical timing effect 
caused by the revolution of the planet around the barycenter.
In our present approach, ${\rm TTV_p}$ is due to the revolution of the photocenter, which 
both combines 
the dynamical and the photometrical properties of the moon and offers a potential to estimate
the mass and the size of the satellite.}

\subsection{		Expressions for the radius and mass of the moon}

\begin{figure}
	\includegraphics[width=\columnwidth,bb=50 50 400 300]{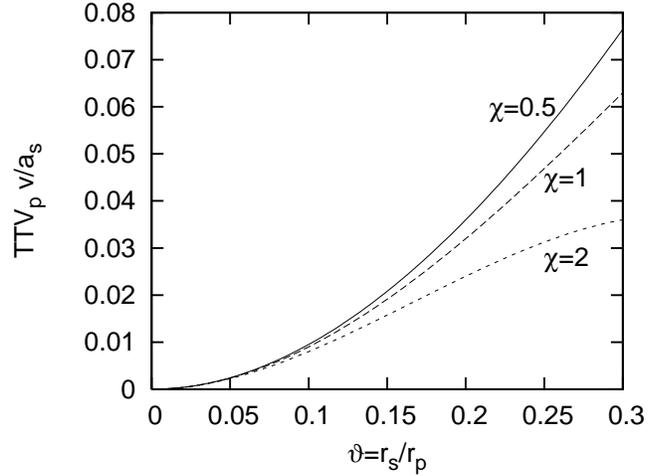}
	\caption{The dependence of ${\rm TTV_p}$ (multiplied by the term $v/a_s$) 
	on the size ratio. The three different curves show three satellites with 
	different density ratios.}
	\label{curve}
\end{figure}

\begin{figure}
	\includegraphics[width=\columnwidth,bb=50 50 400 300]{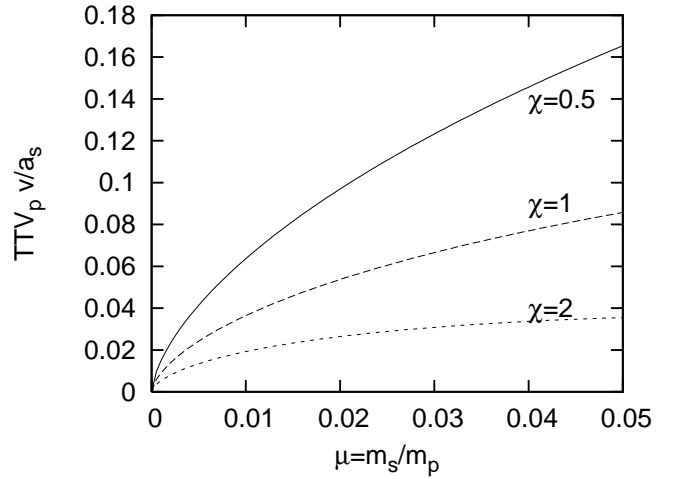}
	\caption{The same as Fig. 3 but for the mass dependence. Note that ${\rm TTV_p}$ depends noticeably on the density in this model,
	which makes the mass determination more ambiguous.}
	\label{curve}
\end{figure}

By expressing the mass ratio $\mu$ through the radii $r_p$, $r_s$ and the densities 
$\rho_p$, $\rho_s$ in Eq. \ref{S1},
we get the formula for the size ratio $\vartheta$:
\begin{equation}
	{\rm TTV_p}= {a_s\over v}{r_s^3 \over r_p^3} \left| {r_p  \over r_s} - {\rho_s \over \rho_p} \right| =
	{a_s\over v}\left| {1\over\vartheta}-\chi\right|\vartheta^3,
	\label{rho1}
\end{equation}
\begin{equation}
	\Delta {\rm TTV_p} = {a_s\over v}\left\{ \left[ 2\vartheta -3\vartheta^2\chi\right] \Delta\vartheta - 
	\vartheta^3 \Delta \chi\right\}.
	\label{drho1}
\end{equation}
where $\chi=\rho_s/\rho_p$ is the ratio of the densities. Then,
$\Delta {\rm TTV_p}$ expresses the error propagation from $\vartheta$ and $\chi$ to ${\rm TTV_p}$ using the total
derivative of Eq. 16. We note that in most cases ${1\over\vartheta}-\chi>0$ and the 
absolute value can be replaced by normal brackets. 
This is because (i) in the case of a giant planet $\vartheta<<1$, $\chi>1$ can be expected, and (ii) in the case of
an Earth-like planet $\vartheta<1$ and $\chi\approx 1$ is plausible.
If $\vartheta$ is known from a direct photometric detection,
 this equation can be used to determine the {\it density ratio. }

From Eq. \ref{S1} we can eliminate the radii using masses and densities. This gives
the formula for determining the mass ratio,
\begin{equation}
	{\rm TTV_p}={a_s\over v} \left| \left( {m_s/\rho_s \over m_p/\rho_p} \right) ^{2\over3} - {m_s\over m_p} \right| = 
	{a_s\over v}\left|\left( {\mu\over\chi}\right) ^{2\over3}-\mu \right|
	\label{rho2}
\end{equation}
\begin{equation}
	\Delta {\rm TTV_p} =  {a_s\over v}\left\{ \left[ {2\over3\chi^{2\over3}\mu^{1\over3}} -1 \right] \Delta\mu - 
	{2\mu^{2\over3} \over 3\chi^{5\over3}} \Delta \chi\right\}.
	\label{drho2}
\end{equation}
The error propagation analysis shows that Eq. \ref{rho1} is quite stable against the ambiguity in $\chi$, while 
Eq. \ref{rho2} is more sensitive to $\chi$. This means that {\it the photometric method is better for a size determination 
than for a mass determination.}
This is also demonstrated in Figs. 3 and 4, which plot ${\rm TTV_p}$ using 
Eqs. \ref{rho1} and \ref{rho2} for different sizes and masses for some discrete
values of $\chi$.

\begin{table*}
\begin{tabular}{lllllll}
 &            $v$ &    $1/3 a_H$ &  ${\rm TTV_p}$ &   ${\rm TTV_p}\cdot v/ a_s$&   $\vartheta={r_s\over r_p}$&     
 $\mu={m_s\over m_p}$ \\
      &            (km/s)&   ($10^3$ km)&  (s)&   &   &     \\
\hline
$\chi=1$&       29.8&          499.6&            $$912&     	$$0.054&	$$0.274&     $$0.020 \\
$\chi=0.605$&   29.8&          499.6&            $$912&     	$$0.054&	$$0.253&     $$0.010 \\
real        &             &         &                 &                &        0.272  &     0.012  \\
\hline
\end{tabular}
\caption{Test analysis of the simulated transit light curve of the Earth-Moon system, using data sets with the noise
and sampling properties of the Kepler mission.} 
\end{table*}

\subsection{$\chi=1$ estimations}

In the case of the known transits, the $\chi$ relative density is unknown. To make
an order-of-magnitude estimate or an upper limit of the size and the mass of a hypothetical
satellite, we have to choose some value for the density ratio, e.g. $\chi=1$. This choice leads to
\begin{equation}
	{\rm TTV_p} \thickapprox  {a_s\over v}\vartheta^2 \left(1-\vartheta\right), 
	\label{chiexp}
\end{equation}
or
\begin{equation}
	{\rm TTV_p}\thickapprox {a_s\over v} \left| \mu^{2/3}-\mu\right|.
	\label{muexp}
\end{equation}  
If we assume that $a_s=a_H/3$, these equations contain only one unknown parameter.
As the size determination is not too density-dependent (Eqs. 17, 19., Figs. 3, 4), one can expect better results using Eq. 
20 for size determination, but Eq. 21 can also be used for a very rough mass estimation. 

\subsection{Testing with the Earth-Moon system}

A simple test of the calculations described above is to examine the Earth-Moon system.
In Szab\'o et al. (2006) the corresponding numerical calculations are described with the real sizes,
orbital periods, and the solar limb darkening
also taken into account. The resulting $\rm TTV_p$ is $15.2$ minutes (normal amplitude). The
corresponding ${\rm TTV_p}v/a_s$ is 0.054, when substituting $a_s=a_H/3$.

First we calculated a $\chi=1$ models leading to the same timing effect. We found that $\vartheta=0.274$,
$\mu=0.020$. The size of this Moon-model agrees well with the real size of the Moon, and 
the mass is about twice the real value. This precision is, however, acceptable
for a trial estimation for the parameters of an exomoon.
However, if we somehow got to know that $\chi=0.605$
in the Earth-Moon system, we would get an acceptably precise result for both parameters as 
$\vartheta=0.253$, $\mu=0.010$ (Table 1, 2nd row). Both values are concordant with
the real parameters within 20\%{}, and suggest the reliability of this method.
We conclude that he determined sizes agree with the real properties, and an order-of-magnitude
estimate for the mass of the Moon is also possible even when no information on the density assumed.

In order to compare the results $\chi=1$ estimations to other possible density ratios, a grid presentation
of the timing effect is plotted in Fig. 5 based on Eqs. 15 and 18. The lines of constant
$\chi$ and of constant $\vartheta$ values are plotted in the $\mu$ vs. ${\rm TTV_p}\cdot v/a_s$ space. 
With this grid one can determine how the mass
and the size estimates of the Moon vary if we have a reliable estimation for $\chi$. Also, this grid may help in
estimating the size and the mass of a possible exomoon in the case of the positive ${\rm TTV_p}$ detections of the future.

\section{Conclusions}

\begin{figure}
	\includegraphics[bb=67 50 382 300, width=\columnwidth]{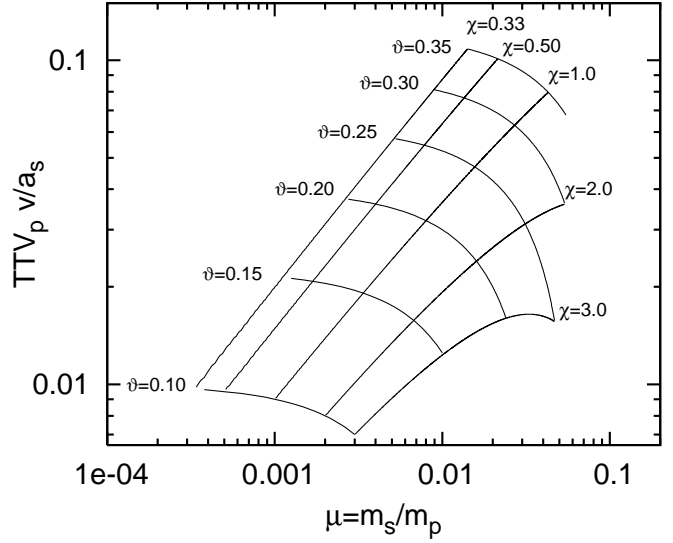}
	\caption{Grid representation of ${\rm TTV_p}$ caused by a satellite with different
	$\chi$ density ratios and $\vartheta$ size ratios.}
\end{figure}

We have presented a novel approach to timing-effect modeling of an exoplanet with a satellite.
This approach offers easy calculations with the light curves and helps estimate the mass and the
size of a possible satellite. The application to high-precision photometric data, such as
is provided by the upcoming satellite missions will lead to estimating the masses and sizes of the moons.
The difference between this approach and the SS99-theory is that here we have
defined the photometric central time of the transit $\tau$, which accounts for
the light curve distortions  due to the planet and the satellite, too. { Although} these seem to be tiny 
corrections, they { have a significant effect in the interpretation of the measurements.} 
An important consequence is that $\rm TTV_p$ cannot exceed
a limit if the mass (or the size) of the satellite is increased more and more. 
To { illustrate this,} let us imagine a ``double planet'', a planet and a satellite having 
exactly the same size and mass. { Because} the positions of this
``planet'' and ``satellite'' are persistently symmetric, one suspects no timing effect to occur. 
Our formulation reproduces this result, if we include $\vartheta=\mu=1$ in Eq. 15. 

In Fig. 5, the grid representation of the ${\rm TTV_p}$ leads to similar conclusions. The constant $\chi$
lines at low $\mu$ values shows that the bigger the size of a moon, the larger the ${\rm TTV_p}$ result. But 
this effect does not increase constantly with size, e.g. a satellite with $\chi=3$ displays maximal ${\rm TTV_p}$
if $\mu=0.034$, and then ${\rm TTV_p}$ starts decreasing with increasing $\mu$. The constant $\vartheta$ lines,
on the other hand, show that ${\rm TTV_p}$ due to a satellite of a fixed size is maximal when the density is low.
This is because the barycenter in this system gets closer to the center of the planet, while the photocenter
remains at its position regardless of $\chi$. 
Therefore the distance between the photocenter and the barycenter increases, 
thus ${\rm TTV_p}$ increases as well, but never exceeds the time between the transit of the photocenter 
and the center of the planet.

The limiting values of ${\rm TTV_p}$ referring to low-density satellites can be calculated from 
Eqs. 16 and 18 generally and from Eqs. 20 and 21 in the $\chi=1$ case, and the highest possible
value of ${\rm TTV_p}$ can be calculated. In this case the timing effect is the largest 
with the value of $v{\rm TTV_p}/a_s\approx0.296$ 
if $\vartheta=2/3$ or, what is the same, $\mu=8/27$. From the viewpoint of the observation, 
too large an upper estimation of ${\rm TTV_p}$ is meaningless whenever it exceeds the above limits.

However, a higher value of ${\rm TTV_p}$ may also be interpreted physically. 
There are more processes that result
in timing effects that can exceed the ${\rm TTV_p}$ caused by an exomoon. A well-known example
is the perturbation of a second planet (Agol \&{} Steffen 2005; Steffen 2006; 
Gillon et al. 2006; Heyl \& Gladman 2006).
Ford \& Gaudi (2006) suggest that the presence of massive exotrojan asteroids can lead to the
shift in the transit central time regarding the time of zero radial velocity difference between the planet
and the barycenter. 
The libration, even on a horseshoe orbit, may result in a timing
effect that can exceed a few minutes.

The results of this paper can be summarized as follows. 
\begin{itemize}
\item{} We gave a theoretical framework for a photometric timing effect caused by moons
of extrasolar planets during transits. We included the photometric processes due to the moon 
via summing the entire photometric signal. This concept offers more parameters to be determined,
such as the sizes and/or the masses of the satellites.  
\item{} The results are supported
by the numerical simulations of Szab\'o et al., and another simple test is described here. With the analysis
of the simulated transits of the Earth-Moon system in front of the Sun, we could almost
reproduce the parameters of { our} Moon exactly.
\item{} We argued that the size of an exomoon is a better predictable parameter than its mass,
and we suggest using this estimation in further analyses.
\end{itemize}

\section{Acknowledgements}
The research was supported by Hungarian OTKA Grant T042509. 
GyMSz was supported by the Magyary Zolt\'an Higher Educational Public
Foundation. { We thank our referee H. Deeg for helping revise this paper.}

\end{document}